# Personalization and Recommendation Technologies for MaaS

Konstantina Arnaoutaki, Efthimios Bothos, Babis Magoutas, Gregoris Mentzas

ICCS, National Technical University of Athens, Athens, Greece.

{konsarna, mpthim, elbabmag, gmentzas}@mail.ntua.gr

## Abstract

Over the last few years, MaaS has been extensively studied and evolved into offering a multitude of mobility services that continuously increase, from alternative car or bike sharing modes to autonomous vehicles, that aspire to become a part of this novel ecosystem. MaaS provides end-users with multimodal, integrated and digital mobility solutions, including a multitude of different choices able to cover users' specific needs in a personalized manner. This practically leads to a range of novel MaaS products, that may have complex structures and the challenge of matching them to user preferences and needs, so that suitable products can be provided to end-users. Moreover, in the everyday use of MaaS, travelers require support to identify routes to reach their destination that adhere to their personal preferences and are aligned to the MaaS product they have purchased. This paper tackles these two user-centric challenges by exploiting state-of-the-art techniques from the field of Personalization and Recommendation systems and integrating them in MaaS platforms and route planning applications.

## 1 Introduction

Mobility as a Service (MaaS) places the user at the core of transport services by offering them tailor made mobility solutions according to their individual needs. In this respect, MaaS users receive customised door-to-door transport services as well as personalized trip planning and payment options (Durand et al., 2018). MaaS operators, the intermediary companies that make agreements with public and private transport operators on a city, intercity or national level, offer subscriptions to bundles of transport services, termed as "MaaS plans" or mobility products (Kamargianni and Matyas, 2017). Access to the transport services is achieved through mobility apps and related back-end platforms that are provided by the MaaS operators and integrate all the available transport services while providing a single point for MaaS package selection, multimodal route planning and payment. Recent research has highlighted the need for the provision of personalization services which should be integrated in MaaS platforms to support user actions, including services that support users to identify MaaS plans that fit their needs when registering in a MaaS service and services that support multimodal route planning during the everyday use of the acquired MaaS plan.

More specifically, in a MaaS environment there can be a multitude of choices in terms of available MaaS plans depending on the specific needs of different travelers and user types. Such choices concern different combinations of available transport services and their levels. For example, MaaS plans can combine and include different levels of public transport, taxi, car sharing, bike sharing, car rental and/or other related services such as parking and e-vehicle charging stations. In such an environment, there is a need to support personalized MaaS plan selection. This is highlighted for example by Polydoropoulou et al. (2018), who surveyed users through focus groups and online questionnaires in the cities of Budapest and Greater Manchester in order to identify features that are required as part of a MaaS solution. In this study, users stated explicitly the need of a self-teaching system that will "learn" the users' needs and suggest the MaaS plans/modes that best matches their lifestyle preferences (e.g. activities/hobbies) and user types (citizens, tourists, children etc.). Furthermore, in a MaaS ecosystem, a multitude of multi modal options is available for commuting from point A to point B and the users need to identify the options that match their preferences, while utilizing their MaaS plans in the best possible way, by considering the current usage of the different modes included in the MaaS plan compared to their allowances. There is therefore a need of providing personalized multimodal route suggestions in the context of MaaS journey planners, by considering user preferences, as well as modal allowances and remaining quotas of the MaaS plan acquired by the user, as also highlighted by Polydoropoulou et al. (2019).

In this work, we describe our approach to provide personalization and recommendation services to MaaS users. More specifically, we have designed and implemented two personalization and recommendation services, the MaaS plans recommender and the MaaS Route recommender that can be integrated in MaaS platforms to address the aforementioned challenges, support related user decisions and improve the MaaS user experience.

The MaaS plans recommender, described in Section 2, exploits methods used in knowledge-based recommender systems, and more specifically in a subcategory of such systems, the constraint-based recommender systems, in order to suggest MaaS plans to end-users/travelers. The basis of our approach is essentially inspired from recommender systems used in domains such as the tourism domain, where the main goal is to optimize the process of selecting a bundle of tourism related products (e.g. activities, places to stay and transport options), the telecommunications domain, where there is the need of supporting the user to select complex mobile phone plans with several features, and the e-commerce domain, where the aim of the bundle is to maximize customers' utility or companies' profits. In our approach, we formulate the MaaS product selection problem as a Constraint Satisfaction Problem (CSP), with the goal of limiting the size of the space that must be searched in order to identify the plans to present to the user.

The MaaS Route Recommender, described in Section 3, supports users in the everyday use of MaaS and more specifically their transportation decisions, by providing a personalised list of multimodal and unimodal routes. Given a list of alternatives route choices for travelling from A to B provided by a routing engine, the MaaS route recommender properly structures the available choices through choice architecture design elements. The choice architecture approach provides proper default options, filters and ranks the route options according to user goals and preferences. Moreover, the recommender service considers optimal use of the MaaS plan the user has subscribed to, as well as the impact on the environment and related long-term effects of potential user choices. Specific goals and preferences of the MaaS operator, such as a preference of a particular mode of transport over another are also considered in the process of structuring the available user choices. To this end and when such a need arises, travellers are nudged towards selecting specific options such as sustainable ones and in the long term change their behaviour and select routes that lead e.g. to reduced emissions in that case. The service can assist urban travellers and commuters to select transportation options that are comfortable, yet satisfying the MaaS operator goals and leading to an optimal use of the MaaS plans.

## 2  MaaS Plans Recommender

## 2.1  Background

It is evident that the selection space of MaaS plans for end users increases according to the available transport services, the combinations of which can generate large choice sets with complex structures. Moreover, despite the fact that travellers make use of mobility services individually and are familiar with them, they are not that familiar with the MaaS concept where mobility services are bundled. Consequently, finding a MaaS product that is aligned to the individual traveller's needs and preferences quickly and accurately is a cognitive task that travellers are not able to manage easily.

The problem of product package selection has been extensively researched within the Recommenders field, primarily seen as the effort to present the correct synthesis of the included parts of a product bundle to a group of users. Schumacher and Rey (2011) demonstrate a study of the different types of recommender systems that can be utilized for dynamic package recommendation in the domain of tourism. Christakopoulou et al. (2016) advocate for the benefits of conversational recommenders that acquire user preferences without much interaction, while underlining on avoiding unpleasant long discussions. Moreover Tumas & Ricci (2009), launch PECITAS, a mobile city transport advisory system, built upon personalized knowledge-based recommenders, that by comparing both travel and user profiles, suggest the top ranked routes to the user.

Constraint technologies, as a type of recommender systems are acknowledged as excessively advantageous for the knowledge acquisition bottleneck, that can be applied easily and have been successfully used for the design of configuration models (Junker and Mailharro, 2003; Mailharro, 1998), whilst a specific implication imposes on constraint satisfaction problems (CSPs). With the term CSP we address a problem, that is constructed out of a finite set of variables, each of which is related with a finite domain, and a set of constraints that reduces the possible values the variables can simultaneously be assigned to. The task is to allocate a value to each variable satisfying all the constraints (Tsang 1993).

## 2.2  Our Approach

Figure 1 provides an overview of our recommendation approach that filters and ranks MaaS plans according to user preferences.

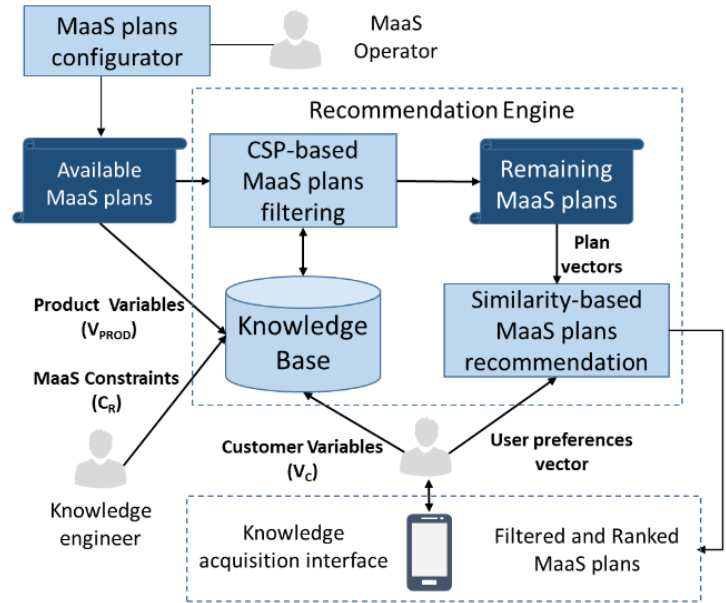

Figure 1: Overview of our knowledge-based CSP and similarity-based MaaS plans recommender.

Since different combinations of offerings and MaaS plans may be available in a city depending on the available transport services and business environment, our approach includes a *MaaS plan configurator* tool that allows MaaS operators to define and configure the MaaS plans to be offered.

The MaaS product selection problem is formulated as a Constraint Satisfaction Problem (CSP), with the goal of limiting the size of the space that must be searched in order to identify the plans to present to the user among those available. The CSP is integrated into a *recommender engine*, where the solution objective is to derive a list of preferred MaaS plans by filtering out plans not satisfying the constraints. The list of remaining MaaS plans is further processed using a weighted similarity function which sorts the results in a ranked list of plans which are aligned with user preferences. In the case that no matching product is found as a solution to the CSP, the similarity-based approach is applied to all available products to rank them based on their similarity to the user profile, under user-provided budget constraints.

Our knowledge-based approach exploits a recommender *knowledge base* that contains explicit rules (*MaaS constraints*) about how to relate user requirements (*customer variables*) with MaaS product features (*product variables*). Such rules are defined by knowledge engineers with knowledge of the field, while user requirements are acquired through questions incorporated into a graphical *knowledge acquisition user interface*.

### 2.2.1 MaaS Knowledge Base and CSP-based MaaS plans filtering

The knowledge base of a constraint-based recommender system can be described through two sets of variables ($V_C$, $V_{PROD}$) and two different sets of constraints ($C_F$, $C_{PROD}$) (Felfernig et al., 2015). These variables and constraints are the vital elements of a constraint satisfaction problem (Tsang, 1993). A solution for a constraint satisfaction problem consists of concrete instantiations of the variables such that all the specified constraints are fulfilled. In correspondence to the Recommender Knowledge Base, under the CSP formalisms stated above, we define the various components of the knowledge base that was developed as the basis for CSP-based MaaS plans filtering as follows:

*MaaS Customer Variables $V_C$*, refer to each user's individual properties. In the domain of MaaS, the frequency of public transport usage is an example for a customer variable and *public transport usage=Every day* represents a concrete customer requirement, indicating a daily use of public transportation services. The most important customer variables along with the questions used to derive them are the following:

- Driving license; derived through the question "Do you hold a full driving license?"
- Public Transport usage; derived through the question "How often do you use public transport?"
- Fare reductions; derived through the question "Are you eligible for any public transport travel fare reductions?"
- CarSharing usage; derived through the question "How often do you use car sharing?"
- Taxi usage; derived through the question "How often do you use Taxi services?"
- BikeSharing usage; derived through the question "How often do you cycle?"

*MaaS Product Variables $V_{PROD}$*, refer to the various attributes of a MaaS plan, including its id, price and the quota per transport mode that is available within the period the plan is valid for (e.g. a month). Examples include the attributes that depict the values of different modes included in the plan (e.g. monthly pass for Public Transport and bike sharing, number of driving hours for car sharing and HUF for taxi services)

*MaaS Products $C_{PROD}$*, refer to the allowed instantiations of product properties, which define the set of available MaaS plans. Indicative examples of MaaS plans are presented in Table 1.

Table 1: Indicative examples of MaaS Plans.

| Id | $V_{PROD}$ | Value |
|---|---|---|
| 1 | Public_transport | 1 monthly pass |
|   | Taxi | 3000 HUF |
|   | Bike_Sharing | 1 monthly pass |
|   | Car_Sharing | 3 driving hour(s) |
|   | Price | 20950 HUF |
| 2 | Public_transport | 1 monthly pass |
|   | Bike_Sharing | 1 monthly pass |
|   | Car_Sharing | 3 driving hours |
|   | Price | 17450 HUF |

*MaaS Constraints $C_F$*, refer to the relationship between Customer and product variables, with the former constraining the values of the latter. Indicative examples of MaaS constraints are provided in Table 2, following an object-oriented annotation language. For example, $C_{F1}$ denotes that MaaS plans which include car sharing are filtered out for users that do not possess a driving license.

Table 2: Indicative MaaS constraints

| Id | $C_F$ |
|---|---|
| $CF_1$ | If user.driving license='No' then MaaS product. CarSharing='0' |
| $CF_2$ | If user.Fare Reductions = 'Yes' then MaaS product. Id='50'or '51' or '52' (special discounted MaaSPlans) |
| $CF_3$ | If user.CarSharing usage ='Every day' MaaS product. CarSharing !=0 driving hours |

Given the user preferences ($V_C$) provided through the aforementioned questions which are embedded in a knowledge acquisition interface, the MaaS product definitions ($C_{PROD}$) and the MaaS constraints ($C_F$), one or more solutions for the constraint satisfaction problem are provided by a CSP solver. The solutions consist of concrete instantiations of the product variables such that all the specified constraints are fulfilled, and correspond to specific MaaS plans that are tailored to the user preferences.

### 2.2.2 Similarity-based Plans Ranking

As already mentioned, our approach includes the calculation of a weighted similarity between a user and MaaS plans, in the direction of ranking the MaaS products that satisfy the constraints (i.e. the output of CSP-based MaaS plans filtering process), on the basis of user preferences for the various modes of transport included in the MaaS plan. Many similarity mechanisms have emerged in Case Based Reasoning (CBR) and data mining research as well as other areas of data analysis. Most of them assess similarity based on feature-value descriptions of cases (e.g. items, users etc.) using similarity metrics that use these feature values. We adopt such an approach that follows the so-called intentional concept description strategy, according to which a concept is defined in terms of its attributes (e.g. a monthly MaaS plan has public transportation, taxi, bike sharing and car sharing usage quotas). This notion of a feature-value representation is underpinned by the idea of a space with cases (e.g. MaaS plans) located relative to each other in this space (Tumas and Ricci, 2009). Similarly, users are represented as a set of feature-value pairs with features representing their preferences for the different modes of transport included in the MaaS plans, in order to allow the calculation of similarity between a user and an item, i.e. a MaaS plan.

Each feature in the representation space is considered to have a different contribution to measuring similarity, i.e. each feature is given a different weight in the user-item similarity calculation. This is because there may be a variance in the importance of each feature for similarity computations, depending on the willingness of each user to include the respective mode in his/her MaaS plan. The higher the willingness to include a mode, the bigger the weight of the respective feature will be. For example, in case a user is more willing to include taxi than bike sharing in a MaaS plan, the taxi feature will be given a bigger weight than the bike sharing one.

The vector representing a user in the X-dimensional feature space (with X denoting the number of distinct modes included in MaaS plans), is instantiated based on user responses to the questions about the frequency of public transport, taxi, bike sharing, and car sharing usage, as described in section 2.2.1. For example, a value of 0 is given to the taxi feature of the user vector, in case the user replies in the relative question, that s/he never uses a taxi service, while the monetary equivalent of 30 taxi rides is given if the user replies in the same question that h/she is using a taxi service once per day. The values for other possible responses will vary between these two extremes with the exception of the case the user replies that s/he is using a taxi service several times per month, where the value is calculated by multiplying the user-provided number with the monetary equivalent of 30 taxi rides. The values for the other features of the user vector are calculated in a similar manner.

The item vectors are instantiated for each MaaS plan based on the values of the features of MaaS Product Variables, i.e. the quota per transport mode that is available within the period the plan is valid for (e.g. a month), as described in section 2.2.1. After the user and MaaS plans vectors have been instantiated for a specific user and a specific list of MaaS plans (the output of the CSP-based filtering), all vectors are normalised and the weighted similarity formula given below is applied to calculate the similarity between the user preferences and all MaaS plans of the list.

$$Similarity(T,S) = 1 - \sqrt{\sum_{i=1}^{F} w_i(T_i - S_i)^2}$$

where $w_i$ is the weight of feature *i*, *T* and *S* are the two input vectors for which similarity should be calculated (i.e. a user and a specific MaaS plan), F is the number of attributes (i.e. features) in each vector (in our case equals the number of distinct modes included in MaaS plans), and *i* is an individual feature from 1 to F. Typically, the weights sum to 1 and are non-negative. The weights are derived from user's response to the following question:

"Please define your willingness to include the following modes of transport in your new MaaS Plan:"

- Public Transport
- Taxi
- Bike Sharing
- Car Sharing

given within a likert-scale 1-5, with 1 indicating "Very much" and 5 "Totally not" option. This question is also embedded in the knowledge acquisition graphical interface depicted in Figure 1. The calculated similarities between the user and the MaaS plans of the list are used to rank the latter and present them to the user in a tabular form in descending order, i.e. the first plan is the most similar to the user preferences and the last one the least similar.

# 3 MaaS Route Recommender
## 3.1 Background

The aim of the route recommendation service is to support users in the everyday use of MaaS and more specifically their transportation decisions, by providing a personalised list of multimodal and unimodal routes. Given a list of alternatives route choices for travelling from A to B provided by a routing engine, the route recommendation service properly structures the available choices through choice architecture design elements. The choice architecture approach provides proper default options as well as filters and ranks the route options according to user goals and preferences while also considering optimal use of the MaaS plan the user has subscribed to, as well as the impact on the environment and related long-term effects of potential user choices. Specific goals and preferences of the MaaS operator, such as a promotion of a particular mode of transport over another are also considered in the process of structuring the available user choices. To this end and when such a need arises, travellers are nudged towards selecting specific options such as sustainable ones and in the long term change their behaviour and select routes that lead e.g. to reduced emissions in that case. The service offers an intelligent decision system, which is tailored for route choice applications and can assist urban travellers and commuters to select transportation options that are comfortable, yet satisfying the MaaS operator goals and leading to an optimal use of the MaaS plans. Our hypothesis is that such applications help urban travellers in the everyday use of MaaS and can result to behavioural changes in the long term.

The use of intelligent decision systems that support sustainable decisions and behavioural change is an emerging field of human computer interaction (Fogg, 2002). Such systems try to 'nudge' users towards decisions that serve their own and the society's at large long-term interests and may take various forms, including gamification systems (Deterding et al., 2011), visual feedback systems (Hargreaves et al., 2010) or systems that properly structure the available choices in decision making contexts. The latter process of structuring the available choices is commonly mentioned as 'choice architecture' (Thaler et al., 2010). It refers to designing and incorporating small features, or nudges, in the choice making process in order to highlight the better choices for the users and assist them to overcome cognitive biases, while not restricting their freedom of choice. The route recommendation service provides filtering and structuring of route choices on behalf of the user whereas it is designed such that it respects user preferences and ensures that the suggested routes are likely to be selected by the user.

## 3.2 Our Approach

The route recommendation service considers a list of alternatives routes for travelling from A to B provided by a routing engine, filters out irrelevant routes, calculates a utility for the remaining routes, and properly structures them on the basis of their utilities, in the sense that the route with the bigger utility is ranked first. The utility of each alternative route represents the degree to which the route satisfies the multiple and in some cases conflicting goals of the route recommendation service. Such conflicting goals may be for example the recommendation of routes that on the one hand match user preferences and the current context and on the other hand contribute to the optimal use of the MaaS plan the user has subscribed to, while satisfying the MaaS operator's goals. The utility calculation is based on:

- generic user preferences
- the current context
- specific user preferences concerning that particular context
- information about a particular route (cost, distance and mode of transport)
- the available quota per transport mode in the user's MaaS package.

Moreover, the service is responsible for estimating a set of context variables that affect which of the routes are relevant for the individual user and the current context. In order to provide route recommendations two main functions are required: i) a filtering function which excludes routes based on the user's profile and other pre-defined restrictions (e.g. very long walking distance), ii) a sorting function which ranks routes in a personalized manner, in order to nudge users towards selecting the routes ranked higher in the list.

We have defined the processing workflow presented in Figure 2 that filters and ranks routes based on information contained in the users' profile and the current context. The output is a formatted route list that makes use of choice architecture design elements in order to highlight the best options.

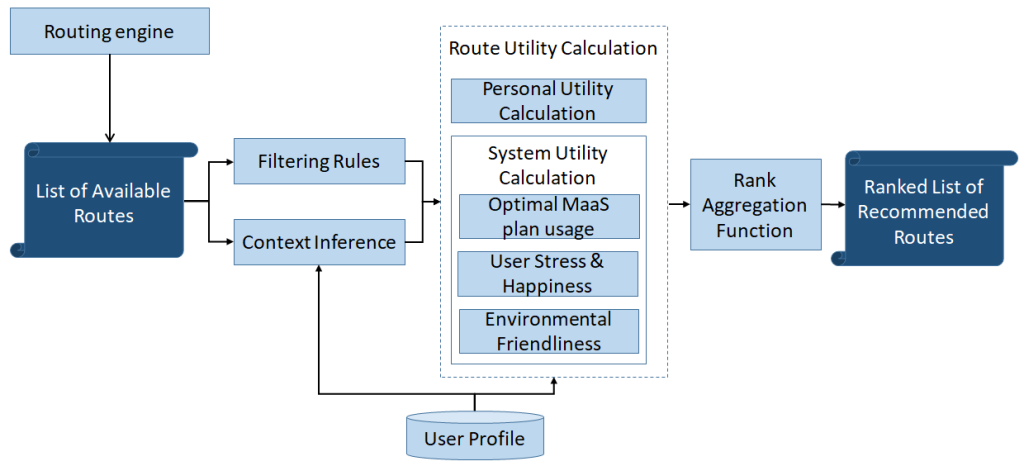

Figure 2: Overview of the processing steps for the route recommendation function

The aim of the filtering rules is to remove route options that do not make sense for the current user. A set of checks has been implemented and each available route undergoes the process of checking for specific route characteristics. In case the system identifies characteristics that are not relevant for the user, the route is removed from the available set.

- For users without a driving license, routes with mode of transport car are excluded.
- For users who cannot cycle, routes with mode of transport bicycle are excluded.
- Routes with long bicycle distances (as defined by the user) are excluded.
- Routes with long walking distances (as defined by the user) are excluded.

### 3.2.1 Context Inference

The route recommendation service leverages context to affect travellers' decisions towards selecting routes that match user preferences, contribute to the optimal use of the MaaS plan the user has subscribed to, while satisfying the MaaS operator's goals. In order to be able to acquire a broad and inclusive understanding of the concept of context in travel choices, we performed an analysis of related studies (Lamsfus et. al, 2015, Gretzel et. al, 2012). Our aim was to collect situational and contextual factors which are relevant to travel behaviour and travel decisions.

The aforementioned analysis resulted in a broad and inclusive understanding of the concept of context in travel choices, based on which we have selected a number of situational and contextual factors which are relevant for a MaaS route recommendation service. The variables are binary, which means that they are activated when the conditions that define them are present and depend on the characteristics of the alternative routes for the current trip, the user's profile and recorded behavior, and the state of the weather. More specifically there are four groups of variables as follows.

1. Based on the users' past behaviour, which are calculated using as input past choices the user has made in the period following the subscription to a MaaS plan and the inferred behaviour as it is logged through the usage of the subscribed MaaS plan. These context variables include: Increased car sharing usage trend; Increased bike sharing usage trend; Increased taxi usage trend; Increased ride sharing usage trend.
These context variables can be activated with the use of sliding-window based functions that analyse the usage of the different modes (car sharing, bike sharing, taxi and ride sharing) in terms of quota that has been used since the start of the window (i.e. the time a user subscribes or renews his/her subscription to a MaaS plan that includes the corresponding modes) and compares it to the uniform consumption of the MaaS plan's available quota for the specific modes. In case the difference exceeds a configurable threshold and the remaining time until the end of the subscription period is below another configurable threshold, the context variable is set to True.
2. Based on trip characteristics, which can be calculated using as input the available routes. These context variables are activated on a per route basis and include the Walking Distance and the Bike Distance.
It means that the total walking or bike distance needed to reach the destination is acceptable by the user. In this case users should be able to configure in their profile their preference with respect to the maximum distance they would be willing to walk or cycle within a multimodal route. In cases when the route walking or bike distance is lower than the threshold set by the user, this context variable is set to True.
3. Based on combination of users' past behaviour and trip characteristics, which can be calculated using as input both past choices the user has made in the period following the subscription to a MaaS plan, and well as the available routes. The following context variable is activated on a per route basis:
Unfamiliar mode or route: Activated when the user is unfamiliar with a route mode or the route itself. User stated preferences, as well as lack of previous user interactions with the MaaS app involving the particular mode or route, are used as measures of route mode and route unfamiliarity, respectively.
4. Based on environmental information. In this case, we make use of information about the environment in which the route recommendation takes place. We define the variable "Nice Weather" which refers to the current status of the weather and is set to True when the temperature level exceeds a certain configurable threshold and the precipitation level in below another configurable threshold.
5. Based on a combination of environmental information and trip characteristics. We make use of information about both the business environment in which the route recommendation takes place and the available routes. The following context variable is activated on a per route basis:
Promoted Mode Route: Activated when the main mode of a route alternative is one that the MaaS operator wants to promote in the current time period. The variable is activated when the total distance that needs to be covered in a route alternative with the mode to be promoted exceeds a configurable threshold.

Promoted Mobility Service Provider Route: This variable is activated when the main mode of a route alternative is one provided by a Mobility Service Provider (MSP) the MaaS operator wants to promote in the current time period.

### 3.2.2 Route Utility Calculation and Ranking

The aim of the Route Utility calculation function is to process the available routes and estimate a personalized utility per route for the specific user in the current context. The utility is used for ranking the routes and presenting them such that routes which adhere to user preferences as well as the current context and contribute to the optimal use of the MaaS plan the user has subscribed to, are ranked higher. The goal is to highlight the routes that lead to the optimal use of the MaaS plan, while respecting user preferences, considering the current context, and increasing their chances of being selected. Eventually, the utility calculation function supports user decisions towards a personalised and context-aware MaaS experience.

The Route Utility calculation function comprises of several sub-functions. In more details the sub-functions provide different views of how the routes should be ordered and presented to the users, which are eventually consolidated in a single ranked list of routes that are communicated to users through the MaaS app. The sub-functions fall under two main views of how the routes should be ordered:

i) The personal user view that considers user preferences and their potential variations in different contexts based on past user interactions with the MaaS4EU application.

ii) The system and context view, which refers to a computational process that leads to the identification of the current context of the user and a user model that infers preferences through the analysis of past behaviour including user trips and selections of routes in a MaaS app. The system view is configured such that it promotes optimal usage of the MaaS plan, a goal reflected in the optimal MaaS plan usage sub-function. Additional goals of the system view that can be optionally activated in the route recommendation service configuration, include the provision of environmentally friendly routes (reflected in the environmental friendliness sub-function), the provision of routes that consider user happiness and stress levels (reflected in the user stress & happiness sub-function), as well as the provision of routes promoting specific transport modes or mobility service providers the MaaS operator wants to promote.

The different route lists are consolidated using the Borda count algorithm and the sum of ranks generated by individual ranking functions to obtain the fused rank (Benediktsson and Kanellopoulos ,1999). Borda Count ranks the routes based on their positions in the basic rankings. If any route has a high ranking in basic rankings it is counted as a high ranking in the final ranking list. The scores of ranking routes in the final ranking list can be calculated as:

$$S_R = \begin{bmatrix} S_{i1} \\ \vdots \\ S_{ij} \\ \vdots \\ S_{kn} \end{bmatrix} \qquad F(S_i) = \sum_{i=1}^{k} S_i$$

Where $S_R$ is a matrix that contains k ranked lists of n alternatives routes in its columns (one for each defined utility function) and $F(S_i)$ is the final score of route i based on its positions in the k ranked lists of routes.

# 4  Conclusions

In this paper, we presented our approach for infusing personalization in MaaS through two recommendation services, namely the MaaS Plans recommender and the MaaS Route recommender. In a MaaS environment there can be a multitude of MaaS plans, that include combinations of transport services, in order to meet the specific needs of different types of travelers. The aim of our service is to support travelers to identify and select MaaS plans and routes and improve their experience. As part of our next steps, we are in the process of evaluating our proposed services in real life conditions where travelers from the cities of Manchester, Budapest and Luxemburg will be using a MaaS app integrating our services for multimodal route planning. Our aim is to test our approach and measure the effectiveness and benefits of MaaS personalization services.

**Acknowledgements**

Research reported in this paper has received funding by the H2020 EC project MaaS4EU (GA no. 723176).